\documentclass[twocolumn,amsmath,amssymb,prl,showpacs]{revtex4}

\usepackage{graphicx}
\usepackage{dcolumn}
\usepackage{bm}

\begin{document}
\bibliographystyle{aip}

\title{Graded anharmonic crystals as genuine thermal diodes: Analytical description of rectification and negative differential thermal resistance}

\author{Emmanuel Pereira}
 \email{emmanuel@fisica.ufmg.br}
\affiliation{Departamento de F\'{\i}sica--ICEx, UFMG, CP 702,
30.161-970 Belo Horizonte MG, Brazil }

\date{\today}

\begin{abstract}
We address the heat flow study starting from microscopic models of matter: we develop an approach and investigate some anharmonic graded mass
crystals, with weak interparticle interactions. We calculate
 the thermal conductivity, and show the existence of rectification and negative differential thermal
resistance. Our formalism allows us to understand the mechanism behind the phenomena, and shows that the properties of graded materials
make them genuine thermal diodes.
\end{abstract}


\maketitle

Many works are devoted to the problem of
understanding the heat flow starting from
microscopic models of matter \cite{LLPD}, and most of them are
 carried out by means of computer simulations, sometimes with
inconclusive results. It creates a demand for analytical studies,
but, since Debye, the microscopic models used to describe heat
conduction are mainly given by systems of anharmonic oscillators,
which involve problems without precise solutions. Anyway,
interesting properties have been discovered and their use
proposed: e.g., the possibility to control the heat flow by using
nano-devices such as thermal diodes, transistors, memories, etc
\cite{Casati, BHu1, BLi1, BLi95, Segal, Chang}.  The basic
structure of these objects, the thermal diode, is a device in
which heat flows preferably in one direction.
 There are analytical attempts to explain this phenomenon  and/or design a diode by using simple
methods \cite{Segal, EM, Ca}, but, again, most of the works are
carried out by means of computer simulations \cite{Casati, BHu1,
BLi1}. A recurrently  used design of diodes is given by the
sequential coupling of chains with different anharmonic potentials
\cite{Casati, BHu1, BLi1}. Although frequently investigated, it is
criticized due to the difficulty to be constructed in practice
\cite{BHu1}. Recently, Chang et al. \cite{Chang} built a diode in
a experimental work by using a different procedure: graded
materials, i.e., nanotubes externally and inhomogeneously
mass-loaded with heavy molecules. Numerical computations
\cite{BLi} also indicate rectification
 in a graded anharmonic system with abnormal conductivity.

An important effect noticed in these studies is the negative differential thermal resistance (NDTR)
\cite{BLi1, HuHe}, a phenomenon where the heat flux decreases as
the applied temperature gradient increases. NDTR is
used to design a thermal diode with a big rectification factor; it
is also crucial for the functioning of some  models
of thermal transistors and logic gates \cite{BLi88}. There
are attempts to explain the origin of NDTR (in systems given by
the coupling of different lattices), see e.g. the
``phenomenological approach''  in ref.\cite{HuPRB10}, but
a general comprehensive understanding of the phenomenon is still lacking.

Hence, considering these central subjects for  the
heat mechanism study,  we address here the following
issues: (i) the development of new  methods of modeling the heat
conduction problem in anharmonic systems;
 (ii) the analytical investigation of the graded mass system
as a reliable candidate for diode, different from that
given by the coupling of different parts, whose rectification
decays with the system size, and that is difficult to be
constructed in practise; (iii) the
understanding of NDTR onset and related properties, in particular, in  a nonlinear
system that is not the coupling of different lattices.

Here, we investigate graded anharmonic crystals with
self-consistent reservoirs, details ahead, and show that graded
materials are perfect candidates for diodes: their rectification
does not decay with size (for certain mass distribution), they
present NDTR, and may be constructed in practice \cite{Chang}. We
recall that rectification is absent in the classical harmonic
version of this model \cite{No}. Our analytical formalism makes
transparent the mechanism behind these phenomena. Rectification
occurs because the total heat flow involves a sum of ``local
conductivities'', see eqs.(\ref{main}, \ref{anhar}), each one
depending on the local temperature (for the anharmonic system, not
for the harmonic one) and also on the masses of neighbor
particles. As we invert the system between two thermal baths, the
distributions of masses and temperatures change in a different
way, leading to a different heat flow - more comments ahead. For
the NDTR onset, we have a competition between gradients of
temperature and mass, see the denominator of the heat flow
expression (\ref{main}, \ref{anhar}). For large gradients and
anharmonicity, there is a change of the dominant term as we
increase the temperature difference, and so, NDTR appears.

Let us introduce the model and our approach. We consider anharmonic
crystals  with stochastic reservoirs at each site. For simplicity, we take $d=1$. We will
work with the ``self-consistent condition'', that means absence of heat flow between each inner
reservoir and its site in the steady state, i.e., the inner reservoirs are not
considered as ``real'' thermal baths as those given by the reservoirs at the boundaries:
they describe only some residual mechanism of
phonon scattering not present in the Hamiltonian. The use of these hybrid models
is recurrent \cite{Lebo1}. Precisely, we take
$N$ oscillators with Hamiltonian
\begin{equation*}
H(q,p) = \sum_{j=1}^{N} [ ~~\frac{1}{2} (\frac{p_j^2}{m_j} + M_j
q_j^2  + \sum_{l\neq j}q_l J_{lj}q_j ) + \lambda\mathcal{P}(q_{j})
] , \label{Hamiltonian}
\end{equation*}
where $M_j >0$, $J_{jl}=J_{lj}$, $\mathcal{P}$ is the anharmonic on-site potential:
$\mathcal{P}(q_j)=q_j^4/4$; with time evolution
\begin{equation}
dq_j = (p_j/m_j) dt , ~~
dp_j = -(\partial H/\partial q_j)dt-\zeta_{j}
p_jdt+\gamma^{1/2}_jdB_j  , \label{eqdynamics}
\end{equation}
where $B_j$ are independent Wiener processes; $\zeta_{j}$ is the coupling between site
$j$ and its reservoir; and $\gamma_j=2\zeta_j m_j T_j$, where
$T_j$ is the temperature of the $j$-th  bath. Here, we will study only nearest-neighbor
interactions.

The energy current inside the system is given by \mbox{$\left<\mathcal{F}_{j\rightarrow}\right>$},
where $\left<\cdot\right>$ means the
expectation with respect to the noise distribution, and
\begin{equation}
\label{flow}
\mathcal{F}_{j\rightarrow} = J_{j,j+1}\left ( q_j-q_{j+1}\right
)\left(\frac{p_j}{2m_j} + \frac{p_{j+1}}{2m_{j+1}}\right),
\end{equation}
precisely, $\mathcal{F}_{j\rightarrow}$ describes the heat flow
from $j$th to $(j+1)$th site; details are found in previous
works \cite{EP1}.

For ease of computation, we map our system on another with
$m_{j}=1$, for all $j$. That is, we make the change of variables:
$Q_j = \sqrt{m_j}q_j$, $P_j =p_{j}/\sqrt{m_j}$, and so, $J$, $M$
and $\lambda$ are replaced by $\tilde{J}_{jk} =
(m_j)^{-1/2}J_{jk}(m_k)^{-1/2}$,
$\tilde{\lambda}_{j}=\lambda_{j}/m_j^2$,
$\tilde{M}_{j}=M_{j}/m_j$. We will drop out the tilde notation in
the unit mass system below, but we make the rescale later to come back
to the general system.

It is also useful to introduce the notation of the phase-space
vector $\varphi=(Q,P)$, with $2N$ coordinates. Then, the dynamics
(\ref{eqdynamics}) becomes
$
\dot{\varphi}=-A\varphi-\lambda\mathcal{P}'(\varphi)+\sigma\eta
$,
where $A=(A^0+\mathcal{J})$ and $\sigma$ are $2N\times 2N$
matrices
\begin{eqnarray*}
A^0=\left (
\begin{array}{cc}
0 & -I \\
\tilde{\mathcal{M}} & \Gamma \end{array} \right ), & \mathcal{J}=\left (
\begin{array}{cc}
0 & 0 \\
J & 0
\end{array}\right ), & \sigma=\left (
\begin{array}{cc}
0 & 0 \\
0 & \sqrt{2\Gamma\mathcal{T}}
\end{array}\right ).\label{defin}
\end{eqnarray*}
$I$ above is the unit $N\times N$ matrix; $J$ is the $N \times N$
matrix for the interparticle interaction $J_{lj}$;  $\tilde{\mathcal{M}}$,$\Gamma$,$\mathcal{T}$ are
diagonal $N \times N$ matrices: $\tilde{\mathcal{M}}_{jl}=M_{j}\delta_{jl}$ ,
$\Gamma_{jl}=\zeta_{j}\delta_{jl}$ , $\mathcal{T}_{jl}=T_j\delta_{jl}$.
$\eta$ are independent white-noises; $\mathcal{P}'(\varphi)$ is a $2N
\times 1$ matrix with $\mathcal{P}'(\varphi)_j=0$ for $ j=1,\ldots,N$
and
$
\mathcal{P}'(\varphi)_i= d\mathcal{P}(\varphi_{i-N})/d\varphi_{i-N}$
 for  $i=N+1,\ldots,2N
$.
In what follows we use the index notation: $i$ for index values in
the set $[N+1,N+2,\ldots,2N]$, $j$ for  values in the set
$[1,2,\ldots,N]$, and $k$ for values in $[1,2,\ldots,2N]$.

In  previous works \cite{EP1} we establish an integral representation
for the correlation functions, and so, for the heat current, of systems with the
stochastic dynamics considered here. It starts with a Gaussian measure, related to the harmonic part of the
interaction. Unfortunately, the analysis
 of the resulting formalism is still very intricate,
in particular, for the case of hard anharmonic potentials. That is, it seems very difficult to reach the anharmonic
behavior starting from perturbations of the harmonic part of the system. Then, in other previous work \cite{EPPhysicaA},
we start an approximative scheme, that we conclude here, within this integral formalism in order to
make it treatable.

Let us describe our approach. Now, we first consider the
equations of dynamics without the interparticle interaction $J$,
but with the anharmonic on-site potential. We do not know a strong
solution for the decoupled anharmonic problem , but we know the
steady distribution: we follow Boltzmann, i.e., our system with $J=0$
involves only noninteracting particles, each one connected to a
thermal bath, and so we have, in the notation $Q,P$,
\begin{eqnarray*}
d\mu_{*}(Q,P) &=& \exp (~ - \sum_{j=1}^{N} H^{(J=0)}_{j}/T_{j}
  ~ ) \prod_{j}dQ_j dP_{j} /{\rm norm.}, \\
  H^{(J=0)}_{j} &=& \left(\frac{1}{2}M_{j}
Q_{j}^{2}
 + \lambda_{j}\mathcal{P}(Q_{j}) + \frac{1}{2}P_{j}^{2}\right).
\end{eqnarray*}
To turn on $J$, we use the Girsanov theorem, which relates the solution of the complete process $\varphi$ (with $J$, the
interparticle interaction) with the previous one $\phi$ (with $J=0$). Precisely, it states that , for $t_{1},\ldots, t_{k}\leq t$,
$
\left < \varphi_{r_{1}}(t_{1})\ldots\varphi_{r_{k}}(t_{k}) \right > =
\int \phi_{r_{1}}(t_{1})\ldots\phi_{r_{k}}(t_{k}) Z(t) d\mu
$,
where $\left< \cdot \right>$ is the expectation for the complete
process $\varphi$, $d\mu$ is the distribution associated to the
expectations of $\phi$ (the decoupled process), and the
``corrective'' factor $Z(t)$ is given by, after manipulations involving It\^o calculus
\cite{EP1,EPPhysicaA},
\begin{widetext}
\begin{eqnarray}
\lefteqn{Z(t) = \exp\left(
-\gamma_{i}^{-1}\phi_{i}(t)\mathcal{J}_{ij}\phi_{j}(t) +
\gamma_{i}^{-1}\phi_{i}(0)\mathcal{J}_{ij}\phi_{j}(0)\right)
 \exp\left( \int_{0}^{t} ds
\gamma_{i}^{-1}\phi_{i}(s)\mathcal{J}_{ij}\phi_{j+N}(s) +\right.} \nonumber \\
&& -
\int_{0}^{t} ds
\phi_{j}(s)\mathcal{J}_{ji}^{\dagger}\gamma_{i}^{-1}A^{0}_{ik}\phi_{k}(s)
 - \left. \int_{0}^{t} ds
\phi_{j}(s)\mathcal{J}_{ji}^{\dagger}\gamma_{i}^{-1}\lambda\mathcal{P}'(\phi)_{i}(s)
-\frac{1}{2} \int_{0}^{t} ds
\phi_{j}(s)\mathcal{J}_{ji}^{\dagger}\gamma_{i}^{-1}\mathcal{J}_{ij}\phi_{j}(s)\right).
\end{eqnarray}
\end{widetext}
We assume the boundary condition $\phi(0)=0$, for simplicity. In the steady state, the
heat flow (\ref{flow}) is related to the expression $
\lim_{t\rightarrow\infty} \left < \varphi_{u}(t)\varphi_{v}(t) -
\varphi_{u-N}(t)\varphi_{v+N}(t) \right > $,  $u>N, v\leq N, $
i.e., $\int \phi_{u}(t)\phi_{v}(t) Z(t) d\mu$, etc. Writing $Z(t)
= \exp(-\int W(\phi(s)) ds$, in a perturbative analysis, we stay
with terms such as $\int [\phi_{u}(t)\phi_{v}(t) W(\phi(s))] ds
d\mu$. But we do not know the distribution $d\mu$, that is very hard to calculate: for the
nonlinear process we know only the steady distribution
$d\mu_{*}$. Then, we introduce an approximative scheme.

First, to relate the fields $\phi(t)$ and $\phi(s)$, we use
the It\^o calculus which establishes that, for functions of
$\phi$: $ \left < f(\phi(t)) \right > = e^{-t\mathcal{H}}
f(\phi(0))$, $\mathcal{H} =  -\frac{1}{2} \gamma_{i}
\nabla^{2}_{i} + \left[ A^{0}\phi +
\lambda\mathcal{P}'(\phi)\right]\cdot\nabla$ , where $\nabla$
means the derivation in relation to $\phi$ (the index $i$, as well
known, takes values in $[N+1,\ldots,2N]$). The difference between
the linear and nonlinear dynamics in the generator of the time
evolution $\mathcal{H}$ above is in the term multiplying the
gradient operator: precisely, instead of $A^{0}\phi$ we have
$(A^{0}+\lambda\mathcal{P}'(\phi)/\phi)\phi$. Thus, to make easier
the calculations, we replace $\phi(t)$ by its average value. Moreover,
 in the exponential relaxation of $\phi$, we still replace
$\mathcal{P}'(\phi)/\phi$ by its average value, more details
ahead. All together means: $ \phi(t) \rightarrow
e^{-(t-s)\mathcal{H}}\phi(s) = e^{-(t-s)\mathcal{A}}\phi(s)$, where
$\mathcal{A}$ is given by $A^{0}$ with $M$ replaced by
$\mathcal{M}\equiv M+\left<\lambda\mathcal{P}'(\phi)/\phi\right>$.
We still have a problem: the computation of
$\int\phi(s)\phi(s)d\mu$ is not possible, since we do not know the
distribution $d\mu$, as said before. Considering that we have an exponential
convergence to the steady state, and so, the main terms involve
$s$ close to $t$, we propose to replace $d\mu$ by $d\mu_{*}$, the
well known steady distribution.

To summarize, our main approximations mean the replacement of
$\phi(t)$ by
 \mbox{$\left<\phi(t)\right>$} and  $d\mu$ by $d\mu_{*}$; after it, the expression for the heat
flow will involve terms such as
\begin{equation*}
\int d\tau \int
d\mu_{*}\left(e^{-\tau\mathcal{A}}\phi\right)
\left(e^{-\tau\mathcal{A}}\phi\right) W(\phi)  ,
\end{equation*}
where the time dependence is carried only by
$\exp(-\tau\mathcal{A})$, where $\tau$ comes from $t-s$, and, as $t\rightarrow\infty$, $\tau\in[0,\infty]$.

In order to teste our approximative scheme, we first turn to the harmonic
self-consistent chains, where rigorous results are known. For a system with
particles with the same mass, and for the case of weak interparticle interactions,
up to first order in $\mathcal{J}$, we have
\begin{eqnarray*}
\lefteqn{\lim_{t\rightarrow\infty}\left <
\varphi_{u}(t)\varphi_{v}(t) \right
>
 =  \lim_{t\rightarrow\infty} \int \phi_{u}(t)\phi_{v}(t) Z(t) d\mu}
\\ &&\simeq \int
\phi_{u}\phi_{v}\left[-\gamma_{i}^{-1}\phi_{i}\mathcal{J}_{ij}\phi_{j}\right]
d\mu_{*} + \int (e^{-\tau A^{0}}\phi)_{u}(e^{-\tau
A^{0}}\phi)_{v} \\
&& \times\left\{ \gamma_{i}^{-1}
\phi_{i}\mathcal{J}_{ij}\phi_{j+N} -
\phi_{j}\mathcal{J}^{\dagger}_{ji}\gamma_{i}^{-1}A^{0}_{ik}\phi_{k}\right\}
d\tau d\mu_{*} ,
\end{eqnarray*}
where $\tau\in[0,\infty]$. After the $\tau$ and $\phi$ integrations, we get
$
 \lim_{t\rightarrow\infty} \mathcal{J}_{uv} \left <
 \varphi_{u}(t) \varphi_{v}(t) \right >  =
 (\mathcal{J}_{uv})^{ 2}(2\zeta M)^{-1}(T_{u} -
 T_{v}) .
$
 That is exactly the same value, considering the lower order in the interparticle
 interaction, of the rigorous computation \cite{EP1}. This expression leads us to the correct thermal
 conductivity. Moreover, for the case of
 a chain with particles with alternate masses (two different values), our scheme also
 works perfectly well: it gives, again, the same value of the rigorous computation.

 Let us, now, analyze our anharmonic crystal. Considering first order in ${J}$,
 with the integration in $\tau$ carried out after using a representation for $e^{-\tau\mathcal{A}}$ \cite{EP1}, we get, for $u>N, v\leq N$,
 \begin{widetext}
 \begin{eqnarray}
 \lefteqn{\left<\phi_{u}\phi_{v}\right> = -(2\zeta_{u}T_{u})^{-1}\mathcal{J}_{uv}\left<\phi_{u}^{2}\phi_{v}^{2}\right> +
 (\mathcal{M}_{v}-\mathcal{M}_{u})(D_{uv})^{-1}\left(\gamma_{u}^{-1} + \gamma_{v}^{-1}\right)\mathcal{J}_{uv}\left<\phi_{u}^{2}\phi_{v+N}^{2}\right> + \nonumber}
  \\
 && \frac{\zeta_{u}+\zeta_{v}}{D_{uv}}\left[\mathcal{M}_{u}\zeta_{v}\gamma_{v}^{-1}\left<\phi_{u-N}^{2}\phi^{2}_{v+N}\right> -
 \mathcal{M}_{v}\zeta_{u}\gamma_{u}^{-1}\left<\phi_{u}^{2}\phi^{2}_{v}\right>\right]\mathcal{J}^{\dagger}_{vu} +
  \frac{\mathcal{M}_{u}}{D_{uv}}\left[(\mathcal{M}_{u}-\mathcal{M}_{v}) + \zeta_{v}(\zeta_{u}+\zeta_{v})\right]\times \nonumber\\
  && \times\left\{(M_{u}\gamma_{u}^{-1} + M_{v}\gamma_{v}^{-1})
 \left<\phi_{u-N}^{2}\phi_{v}^{2}\right>\mathcal{J}_{uv}^{\dagger} +
 \left[\lambda_{u-N}\left<\phi_{u-N}\mathcal{P}'(\phi_{u-N})\phi_{v}^{2}\right>\gamma_{u}^{-1} +
 \lambda_{v}\left<\phi_{u-N}^{2}\mathcal{P}'(\phi_{v})\phi_{v}^{2}\right>\gamma_{v}^{-1}\right]\mathcal{J}_{vu}^{\dagger}\right\} \label{main},
 \end{eqnarray}
 \end{widetext}
where $\mathcal{M}_{u}\equiv\mathcal{M}_{u-N}$,
$D_{uv}= (\mathcal{M}_{u}-\mathcal{M}_{v})^{2} + (\mathcal{M}_{u}\zeta_{v}+\mathcal{M}_{v}\zeta_{u})(\zeta_{u}+\zeta_{v})$.
For $u>N$, $\left<\phi_{u}^{2}\right>=T_{u}$; but the computation of
$\left<\phi_{v}^{2}\right>$, $v\leq N$, is not easy (note that $d\mu_{*}$ is a single variable distribution, and so,
$\left<\phi_{u}^{k}\phi_{v}^{m}\right> = \left<\phi_{u}^{k}\right>\left<\phi_{v}^{m}\right>$). Let us assume some regime
before any approximation: we consider a high anharmonic system, i.e., $\lambda$ large and $M$ small. Thus, we take
$\left<\phi_{v}^{2}\right>= 2c_{1}T_{v}^{1/2}/\lambda_{v}^{1/2}$, $\left<\phi_{v}^{4}\right>= 4c_{2}T_{v}/\lambda_{v}$.
If $M=0$, we would have  $c_{1}\simeq\Gamma(3/4)/\Gamma(1/4)\simeq 1/3$,  $c_{2}\simeq\Gamma(5/4)/\Gamma(1/4)= 1/4$. To determine
the values of $c_{1}$ and $c_{2}$, we turn to the expression of the heat current $\mathcal{F}_{j\rightarrow} =
\mathcal{J}_{uv}(\left<\phi_{u}\phi_{v}\right>-\left<\phi_{u-N}\phi_{v+N}\right>)/2$, with $u-N=j$, $v=j+1$, take all
sites at the same temperature $T$ and find the values such that $\mathcal{F}_{j\rightarrow}=0$. We obtain $c_{2}=1/4$ and
$c_{1}=1/2$. Then, we perform the further computations. For high anharmonicity and very small temperatures, for the dominant term in
$\mathcal{F}_{j\rightarrow}\equiv \mathcal{F}_{j,j+1}$, we obtain,
after the rescaling back to the system with general mass values, i.e., $\lambda_{j}\rightarrow \lambda_{j}/m_{j}^{2}$, etc,
\begin{eqnarray}
\label{anhar}
\lefteqn{\mathcal{F}_{j,j+1} = J^{2}\zeta\left[m_{j}m_{j+1}D_{j,j+1}\right]^{-1}(T_{j} - T_{j+1})} \\
&& \simeq J^{2}\left[\lambda^{1/2}\zeta(m_{j+1}T_{j}^{1/2} +
m_{j}T_{j+1}^{1/2})\right]^{-1}(T_{j} - T_{j+1}) ,\nonumber
\end{eqnarray}
where we take, after the rescale,  uniform potentials and
couplings: $\lambda_{j}=\lambda$, etc. From $\mathcal{F}_{j,j+1}$ above
and the self-consistent condition
$\mathcal{F}=\mathcal{F}_{1,2}=\mathcal{F}_{3,4}=\ldots
=\mathcal{F}_{N-1,N}$, which establishes that the heat
current comes from the first reservoir, passes through the chain
and goes out by the last reservoir, we determine the temperature profile. We
have
\begin{eqnarray*}
\mathcal{F}(m_{2}T_{1}^{1/2}+m_{1}T_{2}^{1/2})/\mathcal{C} &=& T_{1}-T_{2} \\
 &=& \ldots \\
\mathcal{F}(m_{N}T_{N-1}^{1/2}+m_{N-1}T_{N}^{1/2})/\mathcal{C} &=& T_{N-1}-T_{N} ,
\end{eqnarray*}
where $\mathcal{C}=J^{2}/\lambda^{1/2}\zeta$. We sum all the
equations to obtain
\begin{eqnarray*}
\lefteqn{\mathcal{F}\left\{(m_{2}T_{1}^{1/2}+(m_{1}+m_{3})T_{2}^{1/2}+
\ldots\right.} \\
&& \left. + (m_{N-2}+m_{N})T_{N-1}^{1/2}+
m_{N-1}T_{N}^{1/2}\right\}/\mathcal{C} = T_{1}-T_{N},
\end{eqnarray*}
that gives us, from $\mathcal{F}=\mathcal{K}(T_{1}-T_{N})/(N-1)$, an expression for the thermal conductivity $\mathcal{K}$.
The system of equations above may be rewritten as
$$
\frac{T_{1}-T_{2}}{m_{2}T_{1}^{1/2}+m_{1}T_{2}^{1/2}}= \ldots =
\frac{T_{N-1}-T_{N}}{m_{N}T_{N-1}^{1/2}+m_{N-1}T_{N}^{1/2}} .
$$
For the case of particles with the same mass, the equations become $T_{1}^{1/2}-T_{2}^{1/2}= \ldots = T_{N-1}^{1/2}-T_{N}^{1/2}$, that leads
 to a linear profile for $T^{1/2}$, i.e., $T_{k}^{1/2} = T_{1}^{1/2} + [(k-1)/(N-1)](T_{N}^{1/2}-T_{1}^{1/2})$. For a general mass distribution,
 the problem is more complicated: let us examine it in the case of a small temperature
 gradient. We write $T_{1} = T + a_{1}\epsilon$ and $T_{N} = T +
 a_{N}\epsilon$; $T, a_{1}, a_{N}$ and $\epsilon$ given
 ($\epsilon$ small). Then, $T_{j}$ is a function of $\epsilon$,
 with values between $T_{1}$ and $T_{N}$: $T_{j} = T +
 a_{j}\epsilon + \mathcal{O}(\epsilon^{2})$. Let us analyze only
 the first order in $\epsilon$. From the
 equations for the self-consistent condition, we get the solution
 $ a_{j} = a_{1} + (a_{1}-a_{N})\mathcal{S}_{j}/\mathcal{S}_{N}$,
 $\mathcal{S}_{j} = m_{1} + 2m_{2} + \ldots + 2m_{j-1} + m_{j}$. Hence,
 turning to the thermal conductivity formula, after algebraic manipulations,
 we obtain
 $$
 \frac{1}{\mathcal{K}}-\frac{1}{\mathcal{K}'} =
 \frac{\epsilon(a_{1}-a_{N})}{\mathcal{C}(N-1)2T^{1/2}\mathcal{S}_{N}}
 \left[m_{N}^{2}-m_{1}^{2}\right],
 $$
 where $\mathcal{K}'$ is the conductivity for the system with inverted boundary baths.
And so, there is rectification even for a
small gradient of temperature. And more,
if the graded mass grows with $N^{2}$, i.e., $m_{j} = j^{2}\cdot m_{1}$, then the difference above 
does not decay with $N$. By taking $T_{N}>T_{1}$ (i.e., $a_{N}>a_{1}$) and $m_{N}>m_{1}$,
we see that the thermal conductivity is bigger when heat flows from the large to the small
mass, as experimentally observed in a graded system \cite{Chang}.

We stress here that the dependence on temperature for the local
anharmonic conductivity comes from the dynamics:
$\left<\phi(t)\right> \sim e^{(-t\mathcal{A})}\phi(0)$, where
$\mathcal{A}$ depends on $T$ for the anharmonic (not for the
harmonic)  case. The combination of particle masses and
temperatures, and the difference as we invert the chain, lead to
rectification.

Now, we consider the investigation of NDTR. We turn
to eq.(\ref{main}), that is directly related to the heat
flow, and is valid for weak interparticle interaction $J$ in any
regime: low and high anharmonicity, temperature, etc. All the
terms include $D$ in the denominator, except the first one that,
however, may be manipulated and absorbed by the other
terms. Hence, $\mathcal{F}_{j\rightarrow}$ will have $D_{j,j+1}$
in the denominator (see e.g. the first equality in eq.(\ref{anhar}), the expression for
high anharmonic regime), where, we recall, $D_{j,j+1} =
(\mathcal{M}_{j} - \mathcal{M}_{j+1})^{2} + 2\zeta^{2}(\mathcal{M}_{j}
+ \mathcal{M}_{j+1})$, $\mathcal{M}_{j} = M_{j} +
\left<\lambda_{j}\phi_{j}^{2}\right>$, expression determined for a
system with unit masses. For high anharmonicity we have
$\left<\lambda_{j}\phi_{j}^{2}\right> \sim
T_{j}^{1/2}/\lambda_{j}^{1/2}$; and for very low anharmonicity,
$\left<\lambda_{j}\phi_{j}^{2}\right> \sim
\lambda_{j}T_{j}/M_{j}$. Rescaling to get the expression for a
system with different values for the particle masses, and
considering high anharmonicity, just to fix the expression for the
temperature behavior (but the analysis below, adjusting the power of $T$, follows
anywhere), we have, for the first term in $D_{j,j+1}$
\begin{equation*}
(\mathcal{M}_{j}-\mathcal{M}_{j+1})^{2}
 = (c_{j}m_{j+1}  -
 c_{j+1}m_{j})^{2}/(m_{j}m_{j+1})^{2},
\end{equation*}
$c_{j}=[M+\lambda^{1/2}T_{j}^{1/2}]$. The second term in $D$
always increases with $T$, and is subdominant for $\zeta$ small:
precisely, for $\zeta^{2}<\lambda^{1/2}\Delta m_{j}T_{j}^{1/2}/m_{j}m_{j+1}$; note however that it shall dominate for very small $T$, as
assumed in the second part of eq.(\ref{anhar}).
Let analyze the first term, considering
a graded mass chain. For $m_{j+1}>m_{j}$ and $T_{j+1}>T_{j}$,
i.e., gradient of mass and temperature at the same direction, if
$\Delta T_{j} = T_{j+1}-T_{j} \ll \Delta m_{j} = m_{j+1}-m_{j}$,
then $c_{j}m_{j+1}>c_{j+1}m_{j}$; and if $\Delta T_{j} \gg \Delta
m_{j}$, then $c_{j}m_{j+1}<c_{j+1}m_{j}$. Recall that $\Delta
T_{j}$ increases as we increase $T_{j}$, and, of course, $\Delta T_{j}$
depends also on $T_{1}-T_{N}$, the ``total gradient'': $\Delta
T_{j}$ will be very small if $|T_{1}-T_{N}|$ is very small. Hence,
 starting from a very low total
temperature gradient $\Delta T$, as we increase $\Delta T$, then
$(\mathcal{M}_{j}-\mathcal{M}_{j+1})^{2}$ first decreases, but after same point it
becomes an increasing function. That is, $1/D$ first increases and, in sequel,
decreases with $\Delta T$. As we have $\mathcal{F}=\mathcal{F}_{j} \sim D^{-1}\Delta T_{j}$, and
$D$ changes as $\tilde{c}\Delta T_{j}^{1/2}$, with $\tilde{c}$ depending on $\lambda, \Delta m_{j}$,
if $\lambda$ and $\Delta m_{j}$ are not very small, then $\tilde{c}\Delta T_{j}^{1/2}$ dominates
$\Delta T_{j}$ ($T_{j}<1$), and the heat current first increases and then decreases with $\Delta T$.
In other words, we have NDTR.

To conclude, we stress that diodes of graded materials sound to be experimentally
reliable \cite{Chang}, and  ubiquitous structures: our results follow for many other anharmonic potentials as indicated
by the formalism derivation.

Work supported by CNPq and Fapemig (Brazil).

\end{document}